\documentclass[twocolumn,showpacs,preprintnumbers,amsmath,amssymb]{revtex4}
\pdfoutput=1
\usepackage{graphicx}
\usepackage{dcolumn}
\usepackage{bm}
\usepackage{epsfig}
\usepackage{amsfonts,amsmath,amssymb,bm}

\begin{document}

\newcommand{\NB}{N\beta}
\newcommand{\EF}{E_F}

\title{{\it Ab initio} study of electron transport in dry poly(G)-poly(C) A-DNA strands}

\author{C.D.~Pemmaraju$^a$}
\author{I.~Rungger$^a$}
\author{X. Chen$^a$}
\author{A.~R.~Rocha$^b$}
\author{S.~Sanvito$^a$}
\affiliation{a)School of Physics and CRANN, Trinity College, Dublin 2, Ireland\\
b)Centro de Ci\^encias Naturais e Humanas, Universidade Federal do ABC, Santo Andr\'e, S\~ao Paulo, Brazil}

\begin{abstract}
The bias-dependent transport properties of short poly(G)-poly(C) A-DNA strands attached to Au electrodes are investigated with first principles electronic transport methods. By using the non-equilibrium Green's function approach combined with self-interaction corrected density functional theory, we calculate the fully self-consistent coherent $I$-$V$ curve of various double-strand polymeric DNA fragments. We show that electronic wave-function localization, induced either by the native electrical dipole and/or by the electrostatic disorder originating from the first few water solvation layers, drastically suppresses the magnitude of the elastic conductance of A-DNA oligonucleotides. We then argue that electron transport through DNA is the result of sequence-specific short-range tunneling across a few bases combined 
with general diffusive/inelastic processes.
\end{abstract}

\date{\today}

\maketitle
\section{Introduction}
Charge transport in DNA has been the subject of intense research over the past decade \cite{Arc}. This was mainly motivated by potential applications in a variety of technological fields such as nanoelectronics~\cite{Porath0,Ratner1}, gene sequencing~\cite{Marshall, Kelly} and
chemical sensing \cite{DNAChem}.  Furthermore additional interest has stemmed from the still unclear role that DNA electron conduction may play in some biological functionalities such as DNA repair \cite{DNARep}.
Numerous experimental studies on DNA conductivity have yielded a range of conflicting results with insulating~\cite{Braun}, 
semiconducting~\cite{Porath1, Cohen},  metallic~\cite{Xu, Yoo} and even superconducting \cite{Kasumov} behaviours all being observed. 
It has been then rationalized \cite{Endres,Genereux} that the large differences in conductivity observed in different experiments may be due 
to the strong dependence of the charge transport process in DNA on various intrinsic and extrinsic factors. These include DNA conformation, 
base pair sequence, geometric fluctuations, interaction with the solvent and details of the experimental setup such as the molecule-electrode
junction geometry. Nevertheless some consensus is beginning to emerge that charge transfer in DNA is due to a combination of both relatively
short-range coherent tunneling and longer range incoherent diffusion. 

On the theoretical front different approaches have been employed to investigate this complex phenomenon and a number of
comprehensive reviews may be found in the literature \cite{Endres,Cuniberti1,Chakraborty}. {\it Ab initio} methods are capable of 
providing an accurate description of the detailed electronic structure of DNA while treating explicitly solvent and counter ion 
interactions~\cite{depablo, Endres2, Gervasio, Sankey2} as well as the molecule-electrode contact geometry in devices. Classical 
and quantum molecular dynamics simulations have yielded much insight into the changes in DNA electronic structure induced by 
structural fluctuations and solvent dynamics~\cite{Gervasio, Sankey2}. However, {\it ab initio} transport calculations on DNA are 
currently limited to the linear response limit in the static regime because of the large system size and the absence of a tractable 
framework for treating dynamical effects in electron transport from first principles. 

Complementary to {\it ab initio} schemes are model Hamiltonian approaches.  These are well suited and have been employed 
extensively to investigating dynamical effects \cite{Grozema2,Troisi,Hennig,Yamada,Cramer,Grozema,Gutierrez,Woicz,Elstner1} such as 
structural fluctuations and disorder induced by interaction with solvent molecules, both of which are considered crucial to charge 
transfer in DNA. Therefore {\it ab initio} and model Hamiltonian investigations presently complement each other with the parameters 
for model Hamiltonians being extracted from static first principles calculations and efforts to futher develop hybrid approaches 
underway~\cite{Elstner1, Gutierrez2}.

Even in the static limit, the full potential of the {\it ab initio} approach to studying quantum transport in DNA has so far not been realized. 
Most first principles transport calculations to date have been restricted to ideal DNA periodic structures \cite{Sankey1} and studies that 
explicitly consider attachment of DNA oligomers to realistic electrodes have been relatively few \cite{Malla}. Furthermore, transport 
calculations have thus far been carried out only in the zero-bias limit. Interestingly, over the past five years a number of new experiments, which 
can be directly compared to {\it ab initio} calculations, have emerged. These are scanning tunneling spectroscopy (STS) measurements
on single DNA oligomers deposited on Au electrodes and provide a systematic mapping of the electronic density of states (DOS) around 
the Fermi energy. Xu~{\it et al.}~\cite{Endres3} employed ultrahigh vacuum STS to investigate the electronic properties of single thiolated 
12-base-pair poly(GC)-poly(GC) DNA molecules on a Au(111) surface at room temperature. Shapir~{\it et al.}~\cite{Porath2} used transverse 
STS measurements, to map the energy spectrum of poly(G)-poly(C) DNA molecules deposited on gold. Both experiments produced 
semi-conductor-like $I$-$V$ characteristics with a well defined voltage gap. 

In this work, we investigate longitudinal (along the DNA helix axis) electron transport through single poly(G)-poly(C) (pGpC) A-DNA molecules attached to Au(111) electrodes first in the dry and then in moderately hydrated conditions. The homogeneous $\pi$-stacking and relatively small band gap in pGpC DNA in principle provides the best possible conditions for band-like transport. However, already in dry conditions the electronic wave-functions of short pGpC DNA strands (around of 10 base-pairs in length) are strongly localized by an electric dipole along the helical axis. Thus the longitudinal transport in dry DNA is dominated by coherent tunneling through extremely localized states located at various positions along the molecule and the currents are typically tiny. For instance we calculate the entire $I$-$V$ curve of a 6 base-pair pGpC DNA oligomer attached in various ways to the electrodes and show that the currents originating from coherent electron tunneling at voltages of around 1~Volt are several orders of magnitude smaller than those 
observed experimentally through single DNA transport devices. 

Wetting the DNA strand does not improve the coherent transport picture. Water and counter-ions in fact produce a dipolar field, which partially screens the native DNA dipole. This generates additional electrostatic disorder. The DNA molecular orbitals responsible for the electron transport remain largely localized but now their actual spatial distribution and energy position with respect to the electrodes' Fermi level depends on the particular geometrical configuration of the solvent. The associated transmission coefficient as a function on energy is then qualitatively similar to that of the dry case. However, the presence of the solution produces an inhomogeneous level broadening that we calculate to be as large as 1.5~eV and provides an efficient channel for inelastic transport. From this analysis emerges a picture of electron conduction in DNA dominated by coherent electron tunneling on a length-scale of a few base-pairs combined with incoherent diffusion regulated by the solvent. 

The paper is organized as follows. In the next two sections we will introduce our theoretical methods and discuss the basic electronic structure characteristic of pGpC DNA oligomers both in isolation and when attached to Au electrodes. This will set the most appropriate
electronic structure theory for the problem. In particular we will show that correcting the local density approximation (LDA) for electron 
self-interaction \cite{TFSB} is fundamental in this case in order to obtain a realistic molecular level alignment. Then we will move to the
transport calculations by looking first at the zero-bias limit and then to finite bias situations. Finally we will consider the effects of solvation and conclude. 

\section{Methods}
The {\sc Siesta} \cite{Siesta} code, which employs a numerical localized orbital basis set and norm-conserving pseudopotentials 
forms our primary computational platform. In fact most of the density functional total energy calculations and geometry optimizations are 
performed within {\sc Siesta}. In addition the {\sc Gamess}~\cite{Gamess} quantum chemistry package is employed to carry out some 
hybrid-DFT and Hartree-Fock total energy calculations. When {\sc Siesta} is employed, the LDA with the Ceperly-Alder parameterization 
for the correlation energy~\cite{CA} is used as the exchange-correlation functional. Self interaction corrections over the LDA are 
incorporated within the atomic self interaction scheme (ASIC)~\cite{ASIC}. Standard norm-conserving scalar relativistic pseudopotentials 
with the following reference configurations are used for the different atomic species:~H 1$s^1$, C 2$s^2$2$p^2$, N 2$s^2$2$p^3$, 
O 2$s^2$2$p^4$, Na 3$s^1$, P 3$s^2$3$p^3$, S 3$s^2$3$p^4$, and Au 6$s^1$. An optimized double-$\zeta$ basis set is employed 
for all the atomic species except for Au where a single-$\zeta$ basis is used. Sampling in real space is done through a uniform grid 
with an equivalent cutoff energy of 150~Ry. When using Gamess the SBKJC effective core potentials of Stevens {\it et al.}~\cite{sbkjc} are 
employed together with the associated SBKJC basis set. GGA and hybrid-DFT calculations are carried out using the PBE~\cite{PBE} 
and PBE0~\cite{PBE0} exchange correlation functionals respectively.
\begin{figure}[ht]
\centerline{\includegraphics[width=7.9cm,clip=true]{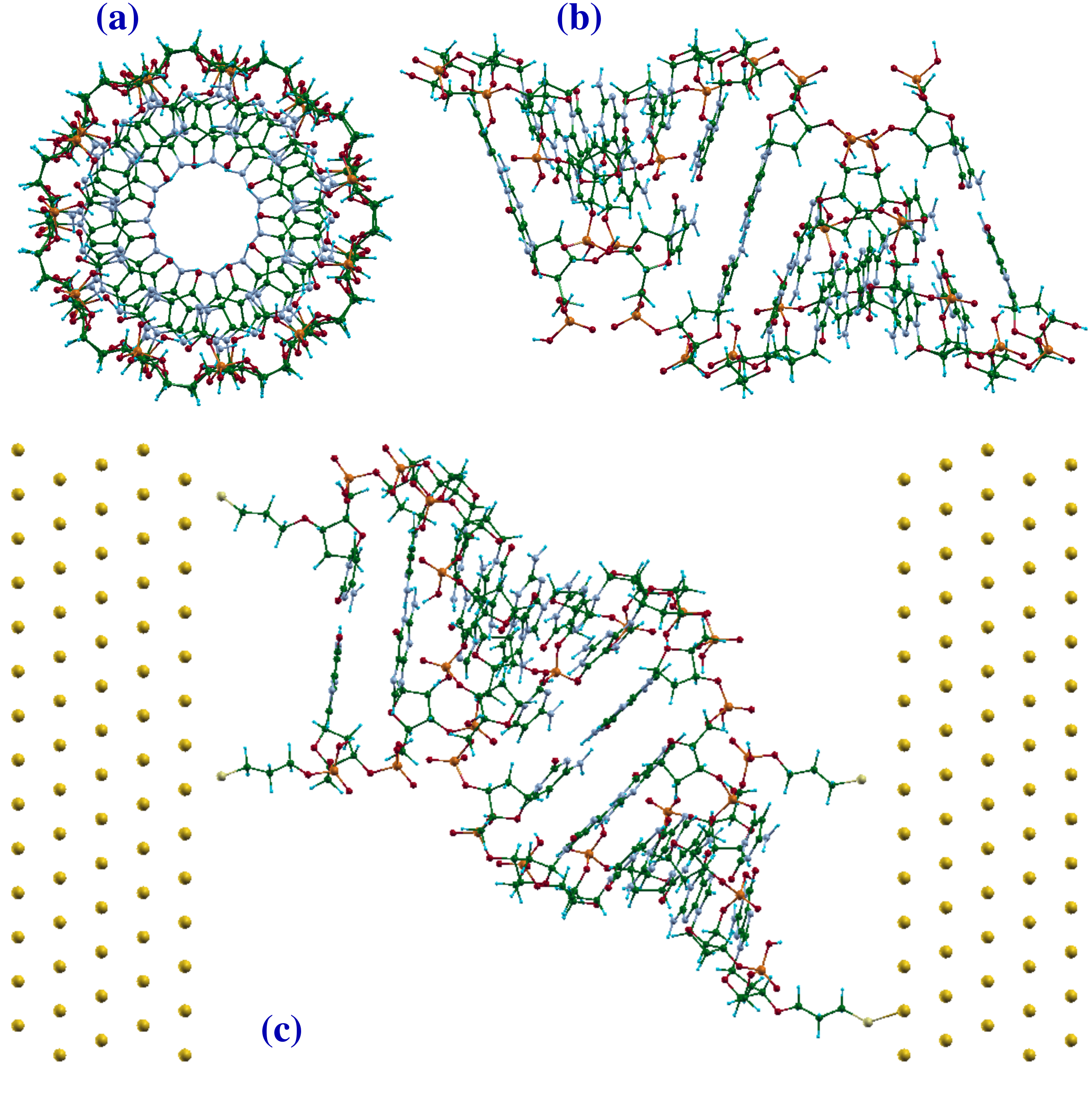}}
\caption{(Color online). Geometric configuration of A-DNA both in its free form and when attached to the electrodes: (a,b) front- and 
side-view respectively of a 11 base-pair poly(G)-poly(C) A-DNA double helix. (c) Two-probe device set up used for transport calculations 
where in the DNA molecule is attached to Au electrodes. Color code: Yellow=Au, Green=C, Red=O, Orange=P, Blue-Gray=N, Cyan=H, Pale-Green=S.}
\label{molpics}
\end{figure}

Transport calculations are performed with the {\sc Smeagol} code \cite{Smeagol1,Smeagol2}, which combines
the non-equilibrium Green's function method (NEGF) \cite{datta,caroli}, with DFT as implemented in {\sc Siesta} \cite{Siesta}. 
The various simulation parameters listed above for {\sc Siesta} are then transferred to the transport calculations. The scattering 
region of the transport simulation setup comprises, in addition to the molecules, five atomic layers of Au(111) included at 
both sides of the DNA. These are sufficient to adequately screen charging at the molecule-Au interface. In order to reduce the system 
size the Au-$5d$ shell is left in the core and only the $6s$ electrons are treated as valence. We use an optimized single-$\zeta$ 
basis for the Au-$6s$ orbitals, specifically tuned to reproduce the correct work function of the Au(111) surface. This is an established and well documented procedure already used with success in the past for describing electron transport across small molecules~\cite{Cormac2}.

The typical size of our simulations and the intrinsic weak coupling between the DNA molecules and the electrodes demand particular care when integrating the Green function yielding the charge density.  Such an integral is split into two parts, one over the complex energy plane and one along the real axis \cite{Smeagol1}. The complex integral is performed over a uniform mesh of 512 imaginary energies, while for the real part we have implemented a new mesh refinement algorithm, necessary to integrate the extremely sharp features of the DOS~\cite{Mn12}. Typically, in order to integrate the non-equilibrium density over a 2 eV window, the final mesh includes 8000-10000 energy points with the denser energy spacing being $\sim 10^{-7}$~eV around the sharpest transmission peaks. Furthermore, in order to boost the convergence for large system sizes, a number of technical solutions have been implemented including a singularity removal procedure to evaluate accurately the leads self-energies \cite{Ivan}. 

\section{DFT electronic structure}
\subsection{Isolated molecules}
The system studied first is an eleven base-pair pGpC DNA molecule with the A-DNA conformation. This constitutes one complete twist of 
the DNA double helix structure (see figure~\ref{molpics}), i.e. in solid state terms it forms the A-DNA one-dimensional unit cell. The molecule, 
consisting of 715 atoms, is $30.5$~\AA\ long and has an average twist angle of $32.72^o$. The negatively charged DNA Sugar-Phosphate 
backbone is neutralized by adding two H ions per base-pair to the PO$_4^-$ groups along the length of the backbone. By imposing periodic 
boundary conditions on this system, an infinitely long DNA chain can be simulated. Alternatively, by suitably pacifying the chain at both ends 
and by relaxing the positions of the atoms at the edges, a finite 11 base-pair oligomer may be modelled. For the periodic structure, we sample 
the Brilluion zone at two $k$-points but even a $\Gamma$-point calculation would be adequate. We present the electronic structure of both 
the periodic and finite chains in order to investigate the differences between the two.

\begin{figure}[ht]
\centerline{\includegraphics[width=7.9cm,clip=true]{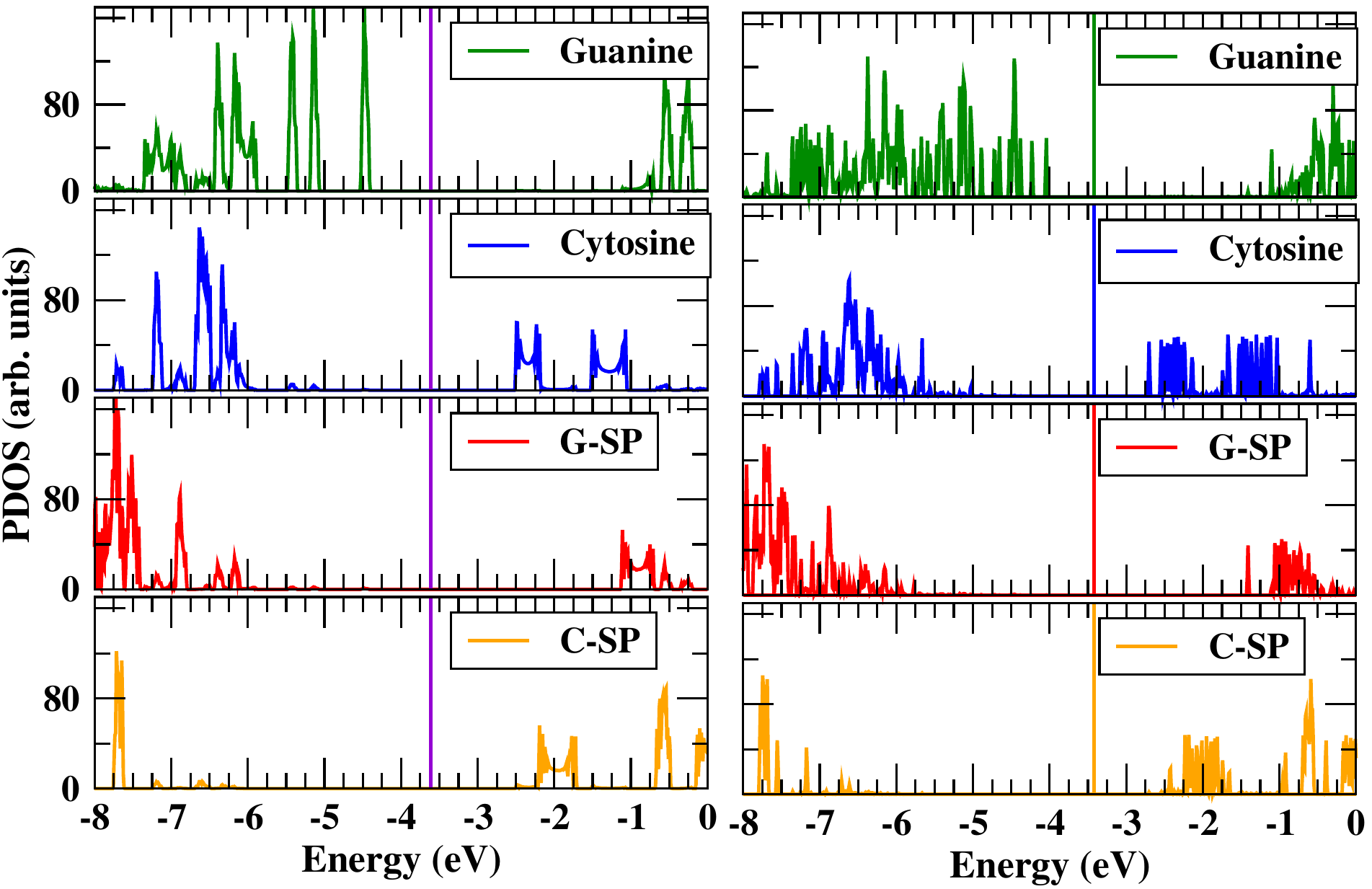}}
\caption{(Color online) Density of states projected onto different constituent parts of poly(G)-poly(C) A-DNA chains: (left panel) projected 
DOS for the infinite periodic DNA structure with 11 base-pairs in the unit cell; (right panel) projected DOS for a finite 11 base-pair oligomer. 
The purple vertical line indicates the Fermi level.}
\label{PDOS_per_fin}
\end{figure}
Figure~\ref{PDOS_per_fin} shows the calculated DOS projectd onto different parts of the DNA chain and displays separately the 
contributions from Guanine, Cytosine and the Sugar-Phosphate (SP) backbone groups corresponding respectively to the Guanine 
(G-SP) and Cytosine (C-SP) side of the helix. In the periodic chain (left panel of figure~\ref{PDOS_per_fin}) a well defined band-like 
DOS appears and the individual bands are well separated from each other.
\begin{figure*}[htbp]
\centerline{\epsfxsize=16.5cm\epsffile{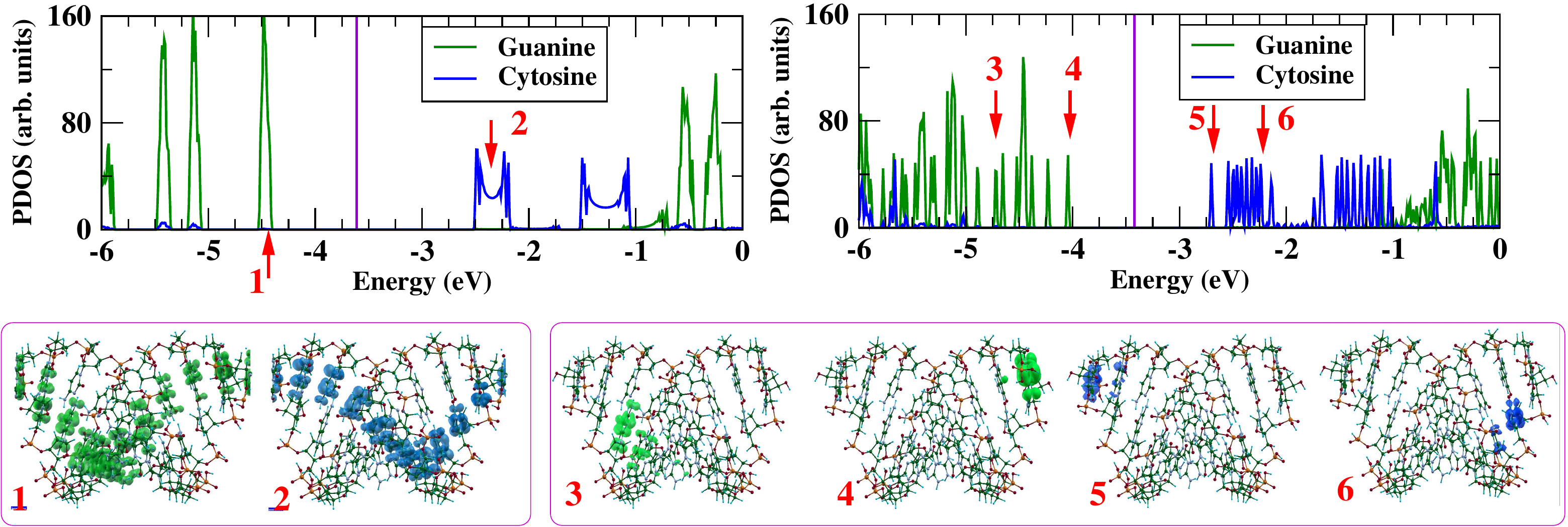}}
\caption{(Color online). Projected DOS (top panels) and charge density distribution (lower panels) for both infinite (left) and finite (right) 
DNA strands. The projected DOS are displayed over an energy window showing the first Guanine and Cytosine levels around the Fermi level.
The arrows and the corresponding numbers mark the various energy levels, whose corresponding charge density distributions are displayed 
in the lower panels. Note the delocalized nature of the HO and LU bands of the infinite chain in comparison with the localized states
of the finite fragments.}
\label{all_pdos_ldos}
\end{figure*}
The highest occupied (HO) band is derived from Guanine and has a rather narrow bandwidth of about 60~meV, while the lowest unoccupied 
(LU) band corresponds to a set of Cytosine states with a relatively larger bandwidth of 300 meV. The system has an overall (LDA) bandgap of 
2~eV. Among the occupied bands, the first three highest ones are derived from orbitals localized on Gunaine and there is little Cytosine or 
backbone related DOS upto 1.5 eV below the HO band. Among the empty states, the first Cytosine-derived band is followed by another 
originating from orbitals localized at the Sugar-Phoshate groups attached to the Cytosine chain. 

When going from the periodic to the finite molecule (right panel of figure~\ref{PDOS_per_fin}), each band splits up into 11 discrete levels 
and the derived energy eigenvalues now spread across a much wider energy range. For instance, individual Guanine-derived eigenstates 
are spread over a 400~meV range on both sides of the 60~meV HO band previously calculated for the periodic case. As a result, the highest 
occupied molecular orbital (HOMO) and the lowest unoccupied molecular orbital (LUMO) are now separated by $\sim$1.25~eV. In order to explore the effects of having finite fragments instead of infinite periodic DNA chains on the electronic density we plot the charge 
density corresponding to different representative eigenstates (see figure~\ref{all_pdos_ldos}). On the one hand the eigenstates from the HO 
and LU bands of the periodic system are seen to spread uniformly across the base pair stacking along the Gunaine and Cytosine chains 
respectively. On the other hand, the eigenstates of the finite molecule are seen to be spatially localized on one or two bases at most. 

This localization of the wavefunctions is the result of the charge dipole along the helical axis created when the periodic DNA chain is 
truncated to form the finite molecule. The dipole points from the 3$^\prime$ end towards the 5$^\prime$ one of the Guanine chain and
it Stark-splits the energy levels corresponding to the various bases. In fact, for the infinite chain the average electrostatic potential at each
individual base is the same for all the bases leading to the formation of extended states albeit with a small bandwidth since the interbase 
hopping is weak. In contrast for the finite chain, the dipole shifts up (down) the average electrostatic potential of the Guanines (Cytosines) 
at the 5$^\prime$ end with respect to those at the the 3$^\prime$. Since such a potential shift is larger than the effective hopping between 
the bases, the electronic wavefunctions become localized. The HOMO is now spatially placed at the 5$^\prime$ end of the Guanine chain and 
states with progressively lower energies are arranged in order from the 5$^\prime$ end towards the 3$^\prime$ one. Similarly, the LUMO is 
now localized at the 5$^\prime$ end of the Cytosine chain. This type of dipole-induced localization is expected to be more pronounced at 
the edges and less in the middle of the molecule especially as the chain length increases. Thus short pGpC DNA fragments in dry conditions, 
even in the absence of other forms of static disorder, support only localized wavefunctions because of the electrostatic dipole along the helical 
axis. 

\subsection{DNA fragments attached to Au electrodes}
We now describe the scattering region setup for a 11 base-pair pGpC strand. DNA strands are connected to the Au(111) surface via 
standard (CH$_2$)$_{3}$-SH thiol linkers. In the reference geometry, the thiol linkers are attached at both the 3$^\prime$ and 5$^\prime$ 
ends of the Guanine and Cytosine chains providing a total of 4 linkages between the molecule and the Au surface. We label this geometry 
G3$^\prime$C3$^\prime$G5$^\prime$C5$^\prime$. Connection to the Au(111) electrodes is made as follows. First, the S atom of one of the 
linker groups is placed at  a hollow site forming an optimal bond length to the surface Au atoms~\cite{Cormac}. The molecule is then re-oriented 
so that the distance from the Au surface of the S atom on the second linker group is optimized. This procedure fixes the orientation of the molecule. 
The second Au electrode is then rigidly shifted to optimize its distance from the two thiol linker  groups on the opposite side of the molecule. 
The device simulation cell includes five Au atomic planes on each side of the molecule. The lateral dimensions of these  are those of a $9\times7$ 
supercell constructed from a rectangular unit cell along the {\it fcc} (111) direction. The unit cell for the electrodes is also a $9\times7$ supercell. 
A lattice constant of $a$~=~4.1812~\AA\ is used for Au. Periodic boundary conditions are enforced in the plane transverse to the transport direction. 
The final simulation cell contains 2017 atoms, of which 1260 are Au atoms of the electrodes. The basis set expands over 5320 local orbitals.
\begin{figure*}[htbp]
\centerline{\epsfxsize=17.5cm\epsffile{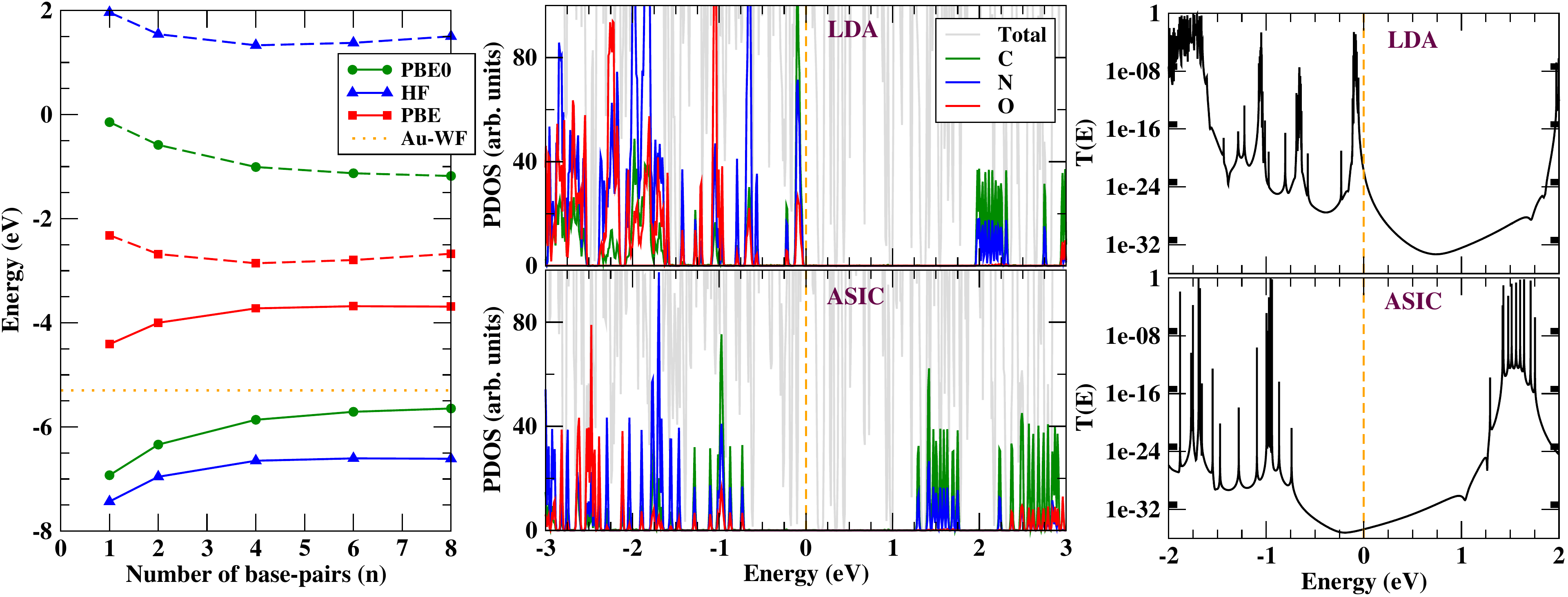}}
\caption{(Color online). DNA electronic structure calculated at various levels of approximation and level alignment between the molecule
and the Au electrodes. In the left panel we show the negative of the ionization potential (solid lines) and of the electron affinities (dashed lines) 
calculated from PBE0-$\Delta$SCF and Hartree-Fock for pGpC DNA strands of different lengths. Also included for comparison are the highest 
occupied and lowest unoccupied eigenvalues from PBE calculations and the Au workfunction. In the middle panel we present the atom projected 
DOS obtained respectively from LDA and ASIC calculations for the 11 base-pair DNA molecules attached to Au electrodes with the reference 
G3$^\prime$C3$^\prime$G5$^\prime$C5$^\prime$ geometry. Finally the right panel displays the zero-bias transmission coefficients as calculated 
with LDA and ASIC for the same system. The Au Fermi level is set to 0~eV.}
\label{g3c3g5c5_ldasic_all}
\end{figure*}

Next we look at the electronic structure of DNA strands attached to Au electrods. In NEGF transport calculations it is fundamental to describe 
accurately the ionization potential (IP) and electron affinity (EA) of the molecule as well as the work function, $W$, of the electrodes. This allows 
one to reproduce the correct alignment of the molecular levels with respect to the electrodes' Fermi level, $E_\mathrm{F}$. In the case of pGpC
this task is complicated by the fact there is no experimental information about the IP and EA of pGpC fragments. Our strategy is then that of 
employing a DFT functional that reproduces well the experimental IP of Guanine and Cytosine in their gas phase and then to calculate that of 
pCpG fragments as a function of the number of base-pairs in the fragment. The parameter-free PBE0 hybrid-DFT functional has been shown 
previously to reproduce molecular excitation energies in good agreement with experiments~\cite{PBE0}. For Guanine (Cytosine) the 
calculated $\mathrm{IP_{cal}}=8.11$~eV ($\mathrm{IP_{cal}}=8.65$~eV) compares well with the experimental one $\mathrm{IP_{exp}}=8.2$~eV 
($\mathrm{IP_{exp}}=8.9$~eV) \cite{Exp_IP}. Here both the IP and EA are calculated with the $\Delta$SCF scheme. The vertical IP of the molecule 
in vacuum is computed as $\mathrm{IP}$=$E(N-1)-E(N)$, which is the energy difference between the singly positively charged and the 
neutral molecule ($N$ is the total number of electrons at neutrality). Similarly, the EA is calculated as $\mathrm{EA}$=$E(N)-E(N+1)$ 
which is the energy difference between the neutral and the singly negatively charged molecule. 

In the first panel of figure~\ref{g3c3g5c5_ldasic_all}, we plot the $\Delta$SCF-calculated negative IP~(green solid line) and the negative 
EA~(green dashed line) for DNA strands ranging from 1 to 8 base-pairs in length. Interestingly we observe that both the quantities present a 
substantial length-dependence. The IP decreases as the number of pairs increases and finally converges towards a value of around 
5.65~eV for the 8 base-pair chain. In contrast the EA has an opposite behavior, it reduces as the number of base-pairs increases and it
approaches the value of 1.2 eV for 8 base-pairs. 

The negatives of the IP and EA set the alignment of the molecular eigenvalues with respect to the vacuum level. Furthermore, the true 
HOMO-LUMO gap of the molecule is simply EA-IP and it is calculated to be to be 4.47~eV.  In figure~\ref{g3c3g5c5_ldasic_all} we 
also plot (blue lines) the removal and addition energies estimated from Hartree-Fock (HF) calculations by using Koopman's theorem 
(i.e. from the position of the single particles eigenvalues). HF calculations usually tend to overestimate the IP and the EA of DNA strands.
This appears to be the case here with the IP converging to around 6.61~eV for the longer chains, which is roughly 1~eV higher than the 
$\Delta$SCF PBE0 estimate. Similarly, also the EA is underestimated with the LUMO eigenvalues coming out positive (i.e. the additional
electron is not bounded). 

Although the IP and EA can be estimated with good accuracy from $\Delta$SCF calculations the corresponding single particle energy levels,
$\epsilon_\mathrm{HOMO}$  and $\epsilon_\mathrm{LUMO}$ respectively, which are the ones that enter in a NEGF calculation, are not readily 
available in DFT. This is because $\epsilon_\mathrm{HOMO}$ obtained from semi-local DFT functionals is typically too high in energy i.e.
it underestimates the IP. Likewise $\epsilon_\mathrm{LUMO}$ is usually too low. Again this is the situation encountered here for 
the PBE-GGA as demonstrated by the results of figure~\ref{g3c3g5c5_ldasic_all} (red lines). Clearly, $\epsilon_\mathrm{HOMO}$ from PBE 
is too high in energy compared to -IP estimated from $\Delta$SCF. The work function of the Au(111) surface is calculated to be $\sim$5.3 eV 
and it is indicated by the orange dotted line in figure~\ref{g3c3g5c5_ldasic_all}. We then need to correct the DFT single particle levels in such a
way for the $\epsilon_\mathrm{HOMO}$ to correspond to the negative of the IP as calculated with the $\Delta$SCF scheme. This is achieved
by using the ASIC method~\cite{ASIC}, a strategy that has already been successfully employed in earlier 
molecular transport calculations~\cite{Cormac2, Cormac3}.

In general, the ASIC scheme uses a scaling parameter, $\alpha$, ranging between 0 and 1~($\alpha=0$ reduces to the LDA, $\alpha=1$ 
corresponds to the full SIC for 1 electron systems). The value of $\alpha$ used is related to the charge screening properties of the material 
under consideration and it is typically $\sim$1.0 for small molecules, around 0.5 for insulating oxides and vanishes for metals~\cite{ASIC}. 
The electronic spectra of numerous wide gap semiconductors have been studied previously using the ASIC scheme with a value of 
$\alpha$=0.5~\cite{ASIC}. For the case of the DNA oligomers, we choose a value of  $\alpha$ that produces an $\epsilon_\mathrm{HOMO}$ 
approximately equal to the negative IP from $\Delta$SCF calculations. For this choice an $\alpha$ in the range of 0.5-0.6 is found to be 
appropriate. 

In figure~\ref{g3c3g5c5_ldasic_all} (middle panel), we present the DOS for the scattering region in the reference geometry calculated 
using both ASIC ($\alpha=0.6$) and LDA (note that LDA and PBE eigenvalues are practically the same). In LDA $\epsilon_\mathrm{HOMO}$
is pinned at the Au $E_\mathrm{F}$. This happens because within LDA/GGA the gas phase $\epsilon_\mathrm{HOMO}$ of the DNA strand lies above $E_\mathrm{F}$ resulting in the creation of a charge dipole at the metal-molecule junction, which acts to align $\epsilon_\mathrm{HOMO}$ with the Au work-function. 
In fact the sharp peak at $E_\mathrm{F}$ seen in the LDA DOS is comprised of 11 nearly degenerate levels corresponding to the 11 base pairs. These states, whose energies spread across a much wider range when the molecule is isolated, are all brought into near resonance because of the fractional charging of the molecule. In contrast ASIC places $E_\mathrm{F}$ in the DNA HOMO-LUMO gap at approximately 0.7~eV from $\epsilon_\mathrm{HOMO}$. Since $\epsilon_\mathrm{HOMO}$ is already below $E_\mathrm{F}$ and the first empty molecular state $\epsilon_\mathrm{LUMO}$ is well above it, no additional charge dipole is created. Also the structure of the 
HOMO manifold is not altered with respect to the isolated molecule. 

Note that although  $\epsilon_\mathrm{HOMO}$ is aligned with the negative IP within ASIC, $\epsilon_\mathrm{LUMO}$ does not correspond to -EA which lies higher. This is formally not a problem as eigenvalues of empty molecular levels need not equal electron addition energies within DFT. However, the fact that the HOMO-LUMO gap is underestimated with respect to the quasiparticle gap within the current approach should be borne in mind while interpreting results from transport calculations. The picture from the DOS is also carried over into the transmission coefficients calculated from LDA and ASIC (right panel of figure~\ref{g3c3g5c5_ldasic_all}). Becaue of the spurious pinning of $\epsilon_\mathrm{HOMO}$ at $E_\mathrm{F}$ LDA predicts a high transmission at zero bias while ASIC predicts a semiconducting system with very little transmission around $\mathrm{E \simeq E_F}$.

\section{Zero bias results}
\begin{figure}[ht]
\centerline{\includegraphics[width=0.5\textwidth,clip=true]{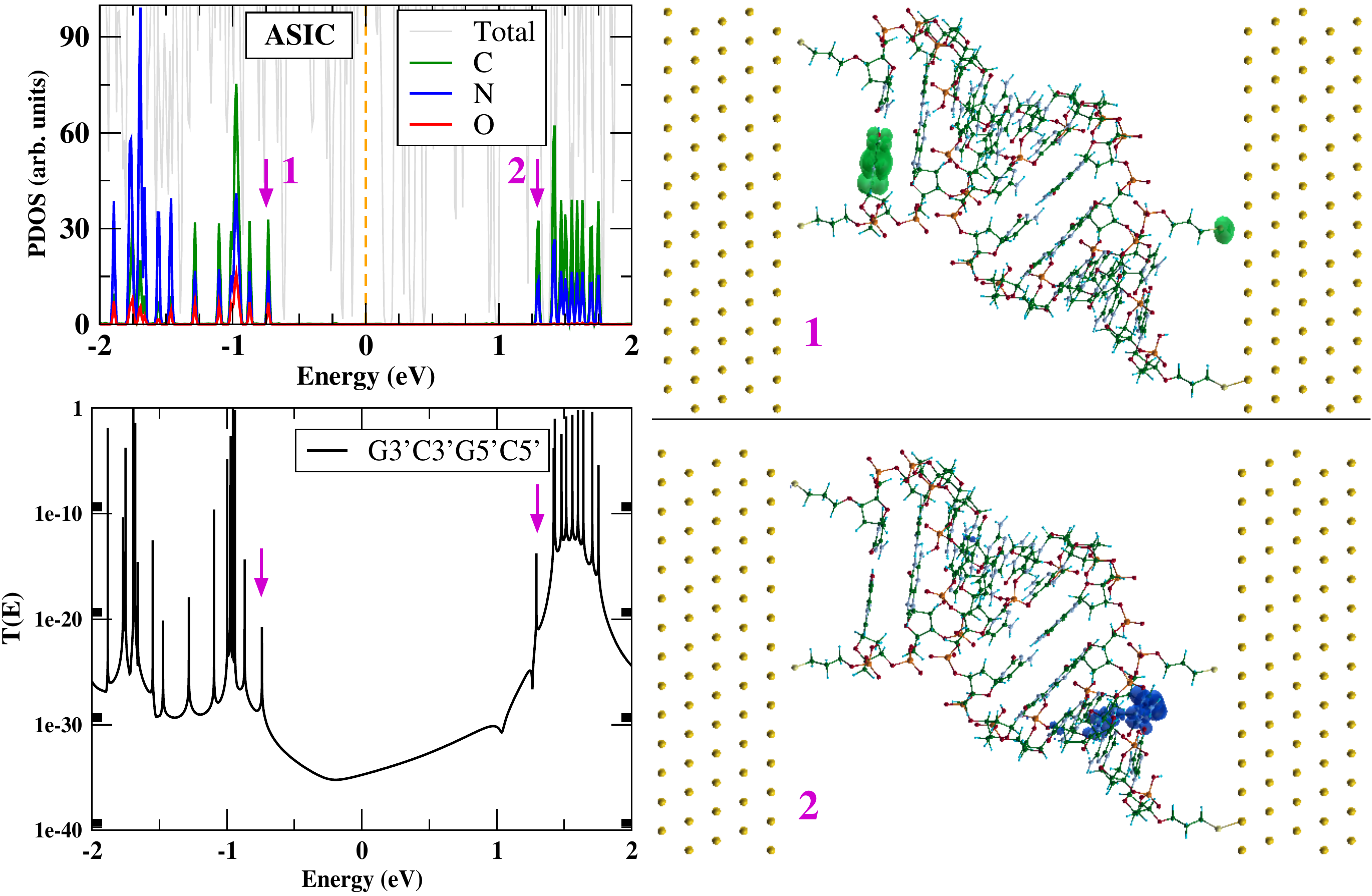}}
\caption{(Color online). Zero-bias transport properties for an 11 base-pair pGpC in the reference G3$^\prime$C3$^\prime$G5$^\prime$C5$^\prime$ geometry. In the top left panel we present the projected density of states. The HOMO and LUMO are indicated by arrows and labelled by 1 and 2 respectively. In the bottom left panel the zero-bias transmission coefficient as a function of energy is presented. In the two right panels we show the charge density distribution corresponding to the HOMO (top) and to the LUMO (bottom). Note that these are localized respectively near the 5$^\prime$ end of the Guanine stack and near the 5$^\prime$ end of the Cytosine stack.}
\label{C4_g3c3g5c5_pldos_trc}
\end{figure}

We next move to analyzing in detail the transport properties of pGpC, by starting from the zero-bias limit. From here on, all the transport results presented as well as the discussion that follows will be based on ASIC calculations. In figure~\ref{C4_g3c3g5c5_pldos_trc} we present the zero-bias transmission coefficient, $T(E)$, as a function of energy, $E$, for the reference geometry (11 base-pairs and connections
to gold through both the strands). The $T(E)$ is typical of a tunneling system and presents rather small transmission values in the gap region away from
both the HOMO and LUMO (note the $T(E)$ is plotted in logarithmic scale). Furthermore, the broadening and maximum height of the different 
transmission peaks, corresponding to the different molecular orbitals, vary significantly across several orders of magnitude. In general, the coupling strength to the electrodes of DNA molecular orbitals that are derived mainly from the Guanine and Cytosine bases is very weak. This is because the bases are well separated from the Au surface by the linking thiol groups as well as the sugar phosphate backbone. As a result the coupling induced broadening of the transmission peaks is very small, typically less than $\sim10^{-5}$ eV for the HO and LU mainfold of states. The resonances corresponding to $\epsilon_\mathrm{HOMO}$ and $\epsilon_\mathrm{LUMO}$ are indicated by arrows and the correspondence 
between the DOS and $T(E)$ is clearly evident. 
\begin{figure*}[htbp]
\centerline{\epsfxsize=16.5cm\epsffile{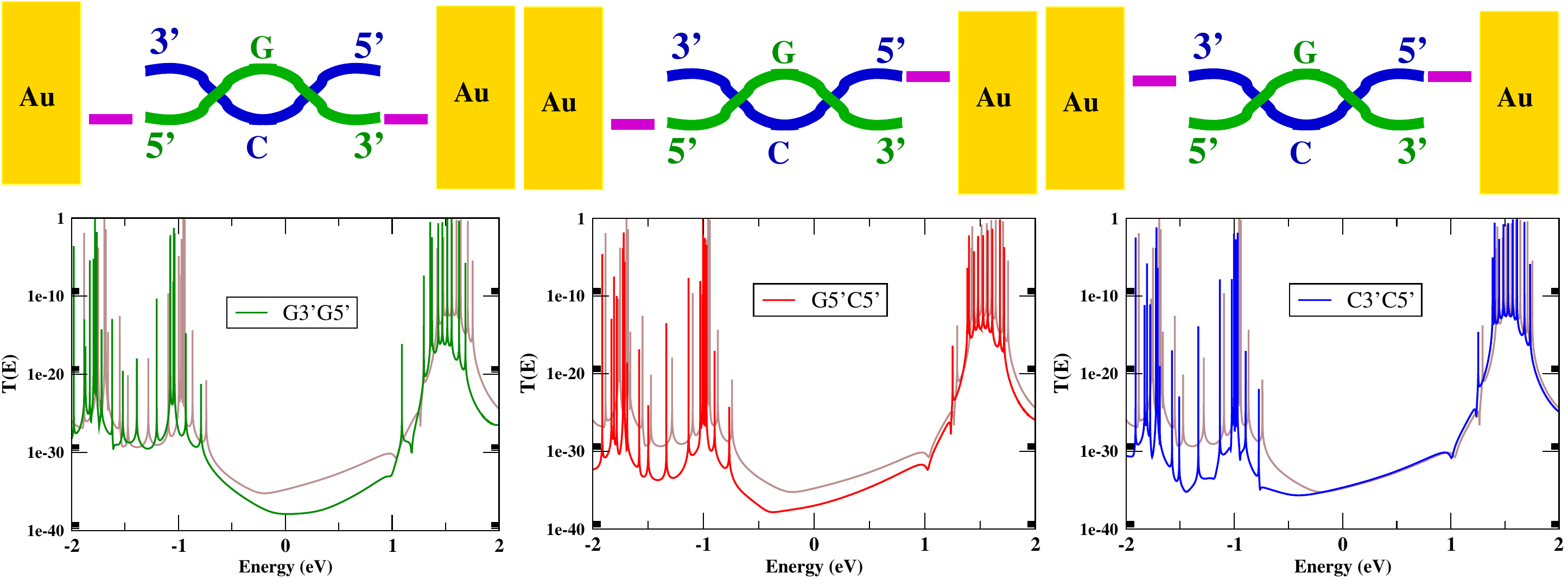}}
\caption{(Color online). Zero-bias transport properties for three different bonding geometries of a pGpC 11 base-pairs DNA strand attached to gold. 
A schematic depicting the geometry of three diffrent molecule-electrode linking configurations is show in the top panels. All the configurations
are obtained from the reference (G3$^\prime$C3$^\prime$G5$^\prime$C5$^\prime$) one, by removing two thiol linkers. The corresponding $T(E)$ for each geometry are presented in the lower
panels. In all the figures the light brown line is the transmission for the reference G3$^\prime$C3$^\prime$G5$^\prime$C5$^\prime$ geometry.}
\label{C4S_3trcs}
\end{figure*}

Depending upon the strength and symmetry of the coupling of the various molecular levels to the electrodes, two types of transmission peaks can 
be identified. Molecular orbitals that are localized on one side of the scattering region couple relatively strongly but asymmetrically to the leads. 
For example, the electronic density for the HOMO level (right panel of figure~\ref{C4_g3c3g5c5_pldos_trc}) is well localized around the 
5$^\prime$ end of the Guanine chain near the left hand-side electrode. Therefore, the HOMO couples far more strongly to the left hand lead than it 
does to the right hand-side one. This asymmetric coupling leads to an attenuated transmission peak at $\epsilon_\mathrm{HOMO}$ \cite{Mn12}. 
Similarly the LUMO charge density is localized on the right hand-side at the 5$^\prime$ end of the Cytosine chain thus couples more strongly to 
the right hand-side lead. In contrast, orbitals that are localized near the middle of the DNA chain couple weakly but more symmetrically to the 
electrodes and thus display corresponding peak heights that are closer to the maximum transmission of 1. As discussed earlier, the eleven 
Guanine derived states in the highest occupied manifold that are sequentially lower in energy are arranged from left to right along the Guanine 
chain. Orbitals corresponding to energy eigenvalues near the center of the HO manifold are localized near the middle of the chain. Furthermore, 
these states are expected to be relatively less localized in space as their corresponding energy eigenvalues are closely packed. This further 
improves the symmetry of their coupling to the leads. A similar analysis also applies to the LUMO manifold. 

Having looked at $T(E)$ for the reference G3$^\prime$C3$^\prime$G5$^\prime$C5$^\prime$ geometry, we now investigate the effect of changing the molecule-electrode linkage geometry. Accordingly, we build three different junctions in which thiol linkers from the G3$^\prime$C3$^\prime$G5$^\prime$C5$^\prime$ configuration are selectively removed and the ends of the DNA chain suitably pacified with H atoms. Furthermore, in removing the alkyl-thiol linker groups, ghost basis orbitals constructed to match the orbitals on the atoms comprising the linker groups are substituted in place of the removed atoms.  Note that in making these new configurations, the position of the DNA chain in space is kept identical to that in the reference setup.  In figure~\ref{C4S_3trcs}, we show schematics of the geometries as well as the corresponding zero bias transmission coefficients. 

\begin{figure*}[htbp]
\centerline{\epsfxsize=17.5cm\epsffile{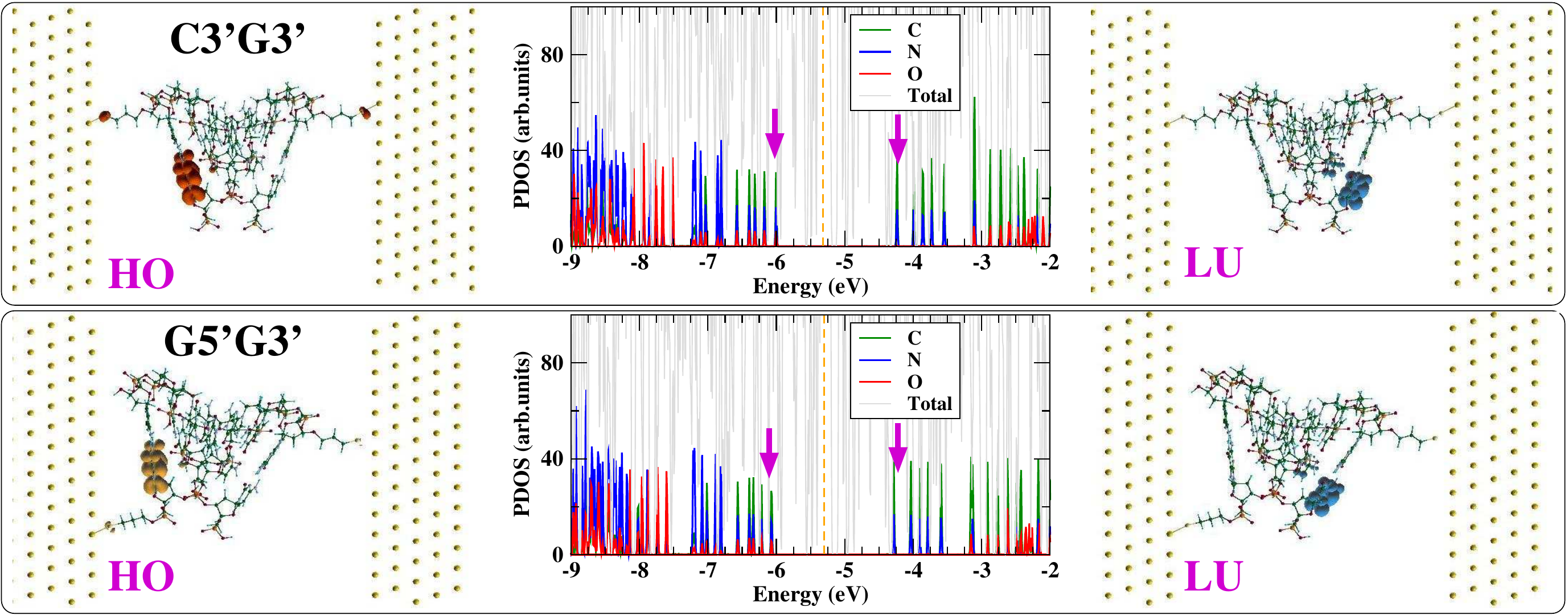}}
\caption{(Color online) Projected density of states and charge density plots for two different device setups comprising 6 base-pair pGpC DNA 
strands attached to Au electrodes. In the top (bottom) left panel we show the charge density corresponding to the HO orbital on the DNA 
molecule connected in the C3$^\prime$G3$^\prime$ (G5$^\prime$G3$^\prime$) geometry. In the central top (bottom) panel the projected 
DOS for the C3$^\prime$G3$^\prime$~(G5$^\prime$G3$^\prime$) setup is displayed. The HO and LU molecular states are indicated 
by arrows. Finally in the top (bottom) right panel we present the charge density corresponding to the LU orbital on the DNA molecule 
connected in the C3$^\prime$G3$^\prime$~(G5$^\prime$G3$^\prime$) geometry.}
\label{6bp_pdos_geom}
\end{figure*}
In the G3$^\prime$G5$^\prime$ configuration, the Guanine chain is attached to Au by thiol groups at both ends, while the Cytosine chain 
is simply pacified and no linkers are connected. Similarly in the C3$^\prime$C5$^\prime$ configuration only the Cytosine strand is attached 
to Au by thiol linkers at both ends while the Guanine chain is left unattached. In both these situations the electrodes are electronically connected only
through one of the two strands of the double helix. In contrast the G5$^\prime$C5$^\prime$ configuration is made by connecting  the 
5$^\prime$ ends of both the Guanine and Cytosine chains to Au via thiol linkages, while leaving the 3$^\prime$ ends free. In this case 
there is no individual strand which is connected at both ends to the electrodes, i.e. electrons must be transferred across the two strands 
in order to sustain a current.

The $T(E)$ plots for these three new geometries are presented in figure~\ref{C4S_3trcs}, where we also reproduce again the zero-bias transmission of the G3$^\prime$C3$^\prime$G5$^\prime$C5$^\prime$ reference geometry (light brown line) for ease of comparison. We note that in the G3$^\prime$G5$^\prime$ geometry $T(E)$ for the Cytosine derived LU manifold of states is suppressed with respect the G3$^\prime$C3$^\prime$G5$^\prime$C5$^\prime$ configuration while the transmission for the Guanine derived HO manifold is virtually unchanged. This is different from the C3$^\prime$C5$^\prime$ junction, where the transmission corresponding to the HO set of states is seen to be similarly suppressed, while that for the LU manifold remains largely unchanged with respect to the reference geometry. Finally, for the G5$^\prime$C5$^\prime$ configuration the transmission is reduced for both the HO and LU manifolds although the largest reduction is for the HO manifold. 

These results show firstly that, even though the thiol linkers are not bonded directly to the Guanine and Cytosine bases but are only attached to the sugar-phosphate backbone, they play a critical role in the transmission. In fact the electron tunneling from the electrodes to the bases is much larger across these organic linkers than through vacuum. Secondly, electron tunneling from molecular orbitals to the electrodes, even for those orbitals that are localized near one end of the DNA strand, effectively occurs through the base-pair stacking. Thus in the 
G5$^\prime$C5$^\prime$ geometry, the broadening of a $T(E)$ peak corresponding to an orbital localized at the 5$^\prime$ end is reduced even though only the thiol linker at the 3$^\prime$ end is removed.

\begin{figure*}[htbp]
\centerline{\epsfxsize=17.5cm\epsffile{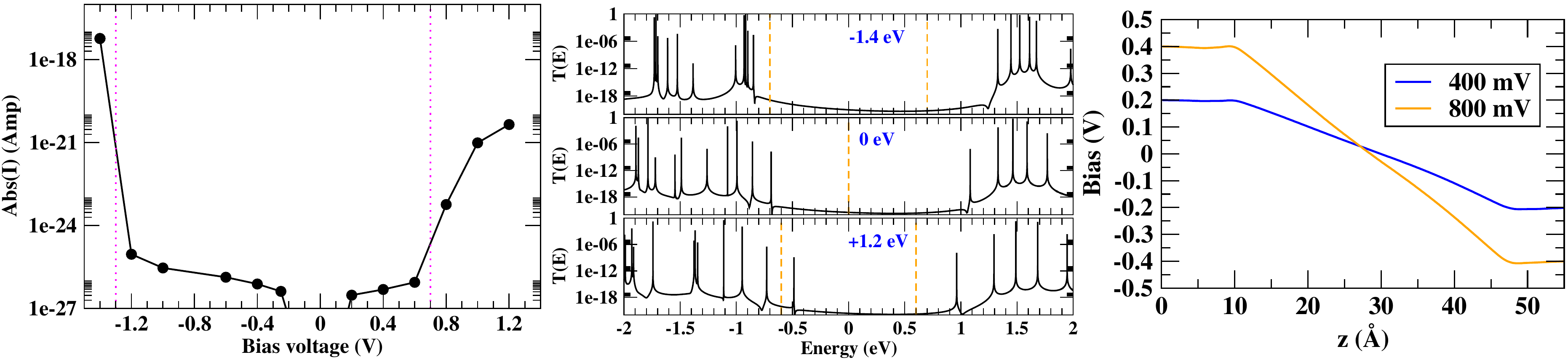}}
\caption{(Color online) Finite bias transport properties of the C3$^\prime$G3$^\prime$ molecule-electrode configuration. In the left panel we
show the $I$-$V$ curve. Note that the current is displayed in logarithmic scale and that for $V<0$ we actually show the negative of the real
current ($-I$). In the central and right panels we present respectively the transmission coefficients as a function of energy at different bias voltages 
and the potential drop across the scattering region. In the $\mathrm{T(E)}$ plots the vertical dashed lines mark the bias windows. }
\label{6bp_pbc_ivtrcpot}
\end{figure*}

\section{Finite bias results}
We now turn our attention to the transport properties of pGpC at finite bias. Since finite bias calculations are considerably more computationally intensive 
than those at zero-bias, we now consider a shorter DNA molecule, namely a 6 base-pair pGpC DNA strand. The construction of the scattering 
region is similar to the one used for the zero bias calculations except for the length of the molecule. The lateral dimension of the electrodes is the 
same as before and periodic boundary conditions are used in the plane. The connection of the molecule to the Au electrodes is once again 
through (CH$_2$)$_{3}$-SH thiol linkers. We construct two different scattering regions differing mainly in the way the molecule is attached to the 
electrodes. In the first configuration labelled  C3$^\prime$G3$^\prime$, the 3$^\prime$ ends of both the Cytosine and Guanine chains are 
connected to Au via thiol linkers while the 5$^\prime$ ends are left free. In the second configuration, labelled  G5$^\prime$G3$^\prime$, 
both the 5$^\prime$ and 3$^\prime$ ends of the Guanine strand are connected to Au while neither end of the Cytosine chain is connected. 
Both these configurations are shown in figure~\ref{6bp_pdos_geom}. 

The atom projected DOS (atom PDOS) for the two systems calculated within ASIC ($\alpha$=0.6) is also shown in Fig.~\ref{6bp_pdos_geom}, 
where both the HOMO and LUMO are marked by arrows. Additionally, the electron density corresponding to the HOMO and LUMO is also plotted. 
The PDOS is qualitatively similar to that of the 11 base-pair case with the nature and relative energy ordering of the different molecular orbitals 
remaining the same. As before, the HOMO charge density is localized at the 5$^\prime$ end of the Guanine chain while the LUMO charge 
density is localized at the 5$^\prime$ end of the Cytosine chain. Thus the strength and the level of symmetry of the coupling to the electrods 
of different orbitals are expected to vary along the base-pair stacks. 

The calculated $I$-$V$ curve for the C3$^\prime$G3$^\prime$ geometry is shown in figure~\ref{6bp_pbc_ivtrcpot}. 
\begin{figure*}[htbp]
\centerline{\epsfxsize=17.5cm\epsffile{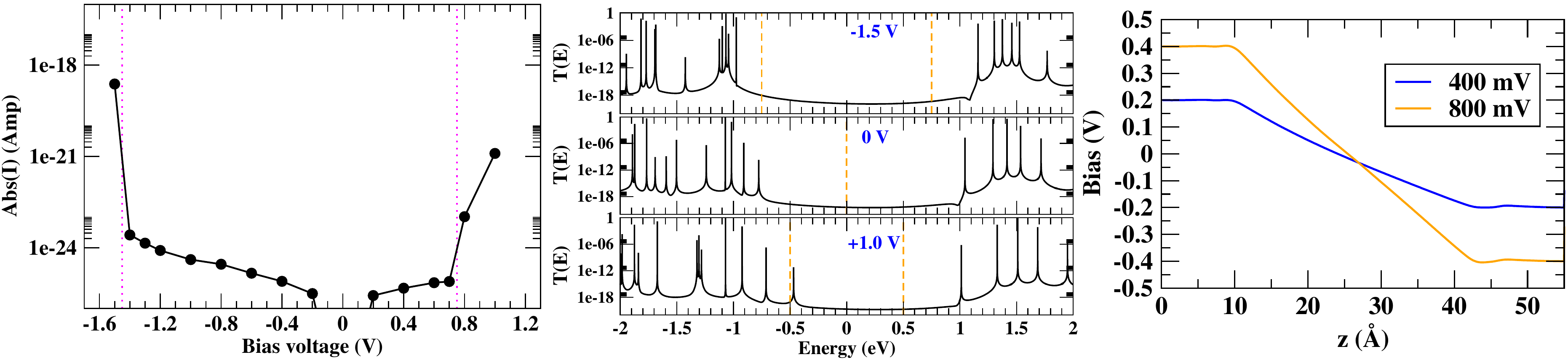}}
\caption{(Color online) Finite bias transport properties of the G5$^\prime$G3$^\prime$ molecule-electrode configuration. In the left panel we
show the $I$-$V$ curve. Note that the current is displayed in logarithmic scale and that for $V<0$ we actually show the negative of the real
current ($-I$). In the central and right panel we present respectively the transmission coefficients as a function of energy at different bias voltages 
and the potential drop across the scattering region. In the $\mathrm{T(E)}$ plots the vertical dashed lines mark the bias windows.}
\label{6bp_g3g5_ivtrcpot}
\end{figure*}
Note that the absolute value of the current is plotted in a logarithmic scale. The magnitude of the current is seen to be very small reaching up
values in the range of 10$^{-21}$-10$^{-17}$~Amperes for bias voltages between 1.2 and 1.4~Volt. These currents are several orders of magnitude smaller than those measured in transport experiments on single DNA oligomers at comparable voltages~\cite{Porath1, Porath2, Endres3}. The potential drop across the device at two different applied voltages is also shown from which it is apparent that the voltage drop occurs almost entirely across the molecule and is flat within the Au layers included in the scattering region. The bias dependent transmission coefficient is also 
plotted in figure~\ref{6bp_pbc_ivtrcpot} for three different bias points. At zero bias the leads' Fermi level lies in the HOMO-LUMO gap of the molecule, where $T(E)$ is extremely small. For positive bias voltages the Fermi level of the left lead ($E\mathrm{_F^L}$) is raised while that of the right lead ($E\mathrm{_F^R}$) is lowered in energy. For negative bias voltages the two Fermi levels are moved in the opposite direction. Since the native dipole on the 6 base-pair DNA strand points from the 3$^\prime$ end towards the 5$^\prime$ end of the Guanine chain, when the 5$^\prime$ end is connected to the left electrode, the external electric field for positive bias acts in the same direction as the native dipole. For negative bias on the other hand the external field opposes the native dipole. As a result, the the transmission peaks for the HO manifold spread out over a larger energy range for positive bias while the same set of states are pushed closer together into near resonance for negative bias. The LU manifold of states also show similar behaviour. Furthermore, for both positive and negative bias, the transmission peak corresponding to the HO level is the first near to the bias window and therefore the voltage gap in this case does not depend upon the position of the LU level. At 1.2~Volt for positive bias for example the transmission peak corresponding to the HOMO is inside the bias window while that of the LUMO is still about 0.4 eV away from it. Similarly for negative bias, the HOMO is within 0.1 eV of the bias window while the LUMO is much farther away. The overall size of the voltage gap for the C3$^\prime$G3$^\prime$  system is seen to be approximately 2~Volt. Such a gap is in good agreement with experimental data.

In figure~\ref{6bp_g3g5_ivtrcpot}, we show the $I$-$V$ curve, $T(E)$ for different bias voltages and the potential drop across the scattering region for the G5$^\prime$G3$^\prime$ configuration. Qualitatively, the results for the G5$^\prime$G3$^\prime$ system are similar to those of the C3$^\prime$G3$^\prime$ one. The current is again very small, in the range of 10$^{-21}$-10$^{-18}$~Amperes for reasonable bias voltages around 1.0-1.5~Volt. Moreover, it is the HOMO level that once again conducts first and sets the size of the voltage gap at about 2.2 eV. Thus in 
spite of marked differences in the molecule-electrode connection geometry, the transport properties of the two device configurations are similar. This is because for the range of voltages investigated here the nature of the $I$-$V$ is determined mainly by a feature common to both geometries, i.e. by the fact that the HOMO is localized at one end of the Guanine chain and it is more strongly coupled to the left lead. 

The extremely small magnitude of the calculated currents at sizeable voltages suggests that even in relatively short DNA strands, the coherent tunneling mechanism is very weak and dynamic or inelastic processes must therefore dominate the charge transfer process.  Additionally, the large contact resistance across the (CH$_2$)$_{3}$-SH linkers in our simulation geometry might also contribute to suppress the current in comparison to experiment. Although in-principle (CH$_2$)$_{3}$-SH linkers are standard in experimental setups~\cite{Porath1, Porath2, Endres3}, one cannot rule out potential imperfections in the junction geometries that allow some parts of the DNA strands to make direct contact with the electrodes thus reducing the contact resistance. Because of their substantial length and large HOMO-LUMO gap, the contact resistance across the (CH$_2$)$_{3}$-SH linkers is expected to be high and indeed the effective coupling to the electrodes of even the DNA bases at the ends of the strand is seen to be rather weak. From the full-width at half-maximum, $\gamma_\mathrm{HOMO}$, of the transmission peak corresponding to the HOMO orbital localized at the  5$^\prime$ end of the Guanine stack, it is possible to make an estimation of the saturation elastic current~\cite{Mn12} for a junction with just one base-pair directly attached to Au electrodes across (CH$_2$)$_{3}$-SH linkers, assuming that the coupling to both electrodes is equal to  $\gamma_\mathrm{HOMO}$. We find $\gamma_\mathrm{HOMO}$ to be $\sim$3$\times$10$^{-6}$eV which yields a value of $\sim$0.3 nA for the saturation elastic current~\cite{Mn12} through such a junction with just one base pair contacted via (CH$_2$)$_{3}$-SH linkers. Thus solvent mediated in-elastic effects might already play a role in the process of electron hopping between the electrodes and the terminal bases of DNA strands.

\section{Solvent effects}
It is well known that the conductivity of DNA depends strongly on environmental effects, in particular on the presence of a solvent and 
additionally counter-ions. The solvation state can influence the electronic structure of DNA in several ways. For instance the confirmation
assumed by DNA double helices is itself determined by the extent to which the molecules are surrounded by water and by the energy 
associated with hydration. Additionally, as shown recently by Rungger {\it et al.}~\cite{Ivan_bdt}, in molecular junctions, solvation by water 
effectively produces electrostatic gating which can change the average alignment of the molecular energy levels with respect to the leads 
$E_\mathrm{F}$. Finally, around room temperature the thermal motion of water molecules can produce rapidly fluctuating local dipolar 
fields that in turn lead to fluctuations in the energy level positions of individual molecular orbitals on the DNA. Such fluctuation may play a 
crucial role in charge transport processes through DNA chains~\cite{Basko}.  
\begin{figure}[htbp]
\centerline{\epsfxsize=0.5\textwidth \epsffile{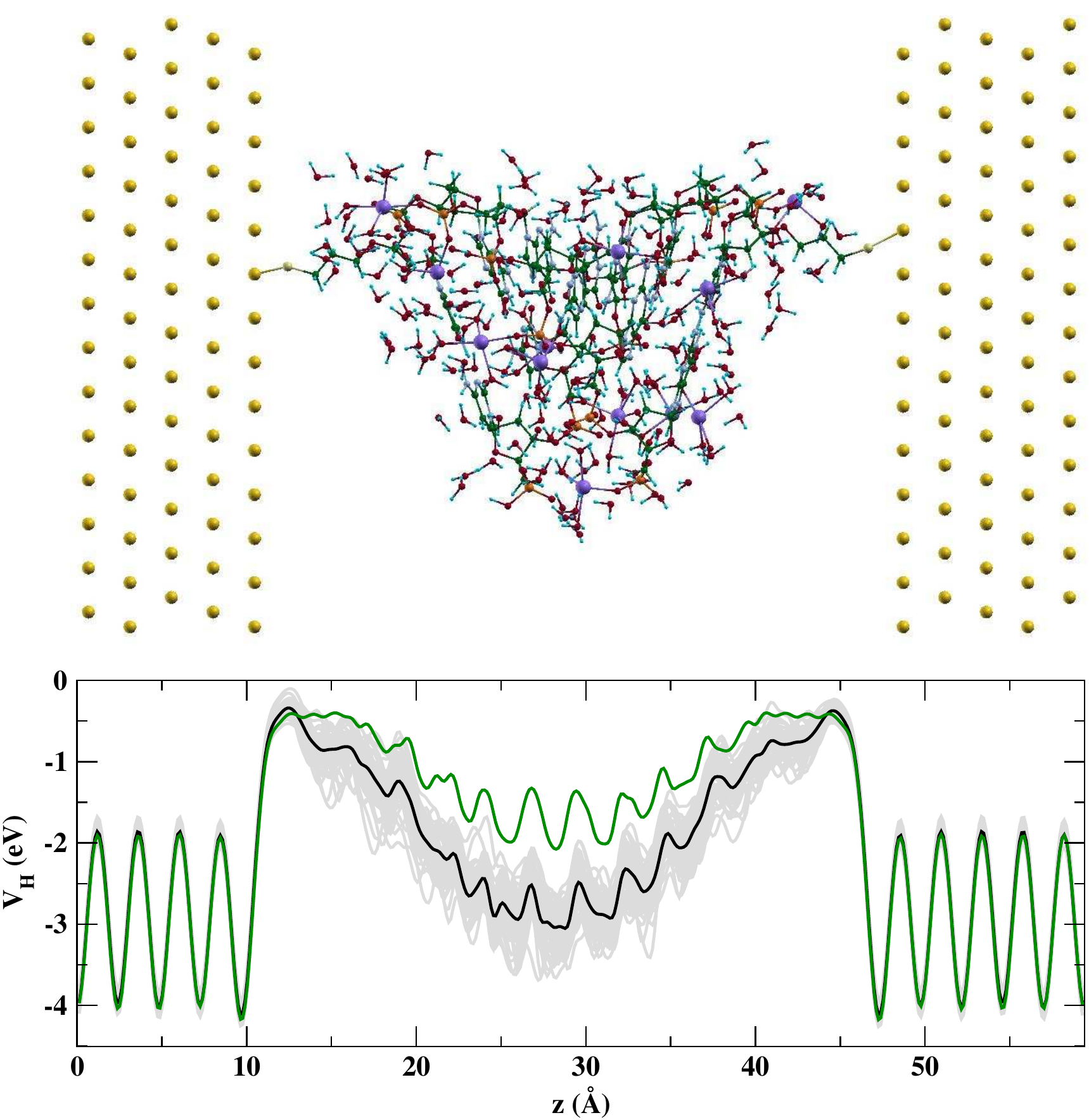}}
\caption{(Color online) A 6 base-pair pGpC molecule in wet conditions. In (a) we show the scattering region setup. This consists of a 6 base-pair pGpC A-DNA strand surrounded by a few layers of solvent including water molecules and Na$^+$ counterions. In (b) we show the planar 
average of the electrostatic potential across the system at zero-bias. The green line shows the electrostatic potential in the absence of the solvent 
while the black line shows the potential averaged over 100 time snapshots of the system taken from the classical MD simulations. The gray 
lines show the electrostatic potential for the individual snapshots.}
\label{6bp_h2o_scatreg}
\end{figure}
\begin{figure}[htbp]
\centerline{\epsfxsize=0.5\textwidth \epsffile{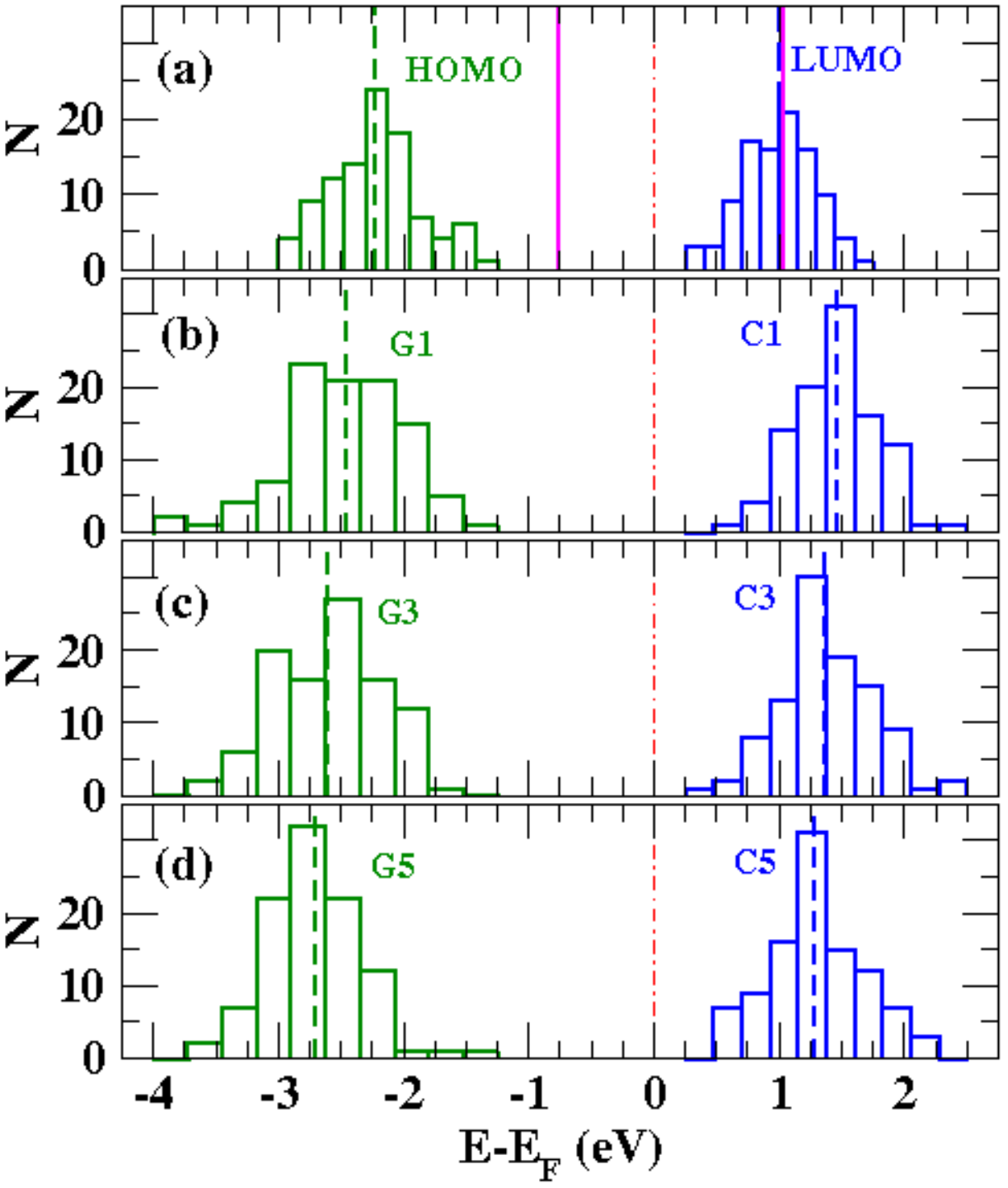}}
\caption{(Color online) Histograms showing the number of times, N, a certain molecular energy level is located within the energy interval given by the position and bin width of the corresponding histogram bar. The purple lines in panel (a) represent the HOMO and LUMO levels of the DNA strand in the absence of the solvent. The dashed lines represent the averages taken over the histograms. Panels (b)-(d) show the distribution of the highest (lowest) occupied (un-occupied) molecular levels localized on individual Guanine (Cytosine) bases. The Guanine bases labelled G1, G2 etc are arranged from left to right with Guanine G1 at the 5$^\prime$ end and G6 at the 3$^\prime$ end.}
\label{HOLU_hist}
\end{figure}

In most STS-based experimental studies on charge transport through DNA~\cite{Endres3, Porath2} the double stranded oligonucleotides are assembled in solution before undergoing a drying process aimed at removing the solvent molecules. However, the exact solvation state of the DNA molecules resulting from such a process is usually unknown and it is likely that a thin solvation layer still surrounds the DNA with the polar water molecules in the solvent hydrating both the negatively charged phosphate backbone of the DNA as well as any counter-ions that may be present. Therefore, in the context of studying electron transport through A-DNA attached to gold electrodes, we also investigate the effects of a thin layer of solvent consisting of water molecules and  Na$^{+}$  counter-ions. We note that nominally wet DNA in the B and Z conformations has been studied  previously by several theoretical groups~\cite{Endres, Gervasio, Malla, Mantz} both within a static picture as well as including the dynamics of water molecules.  In particular, Mantz \textit{et. al}~\cite{Mantz} studied the interplay between solvent geometry and hole charge localization in Adenine-Thymine bridges using QM-MM simulations demonstrating the gating effect due to solvent rearrangements on the pico-second time scale.

Here we are mainly interested in the differences in the electronic structure between the dry and solvated A-DNA strands as calculated within the ASIC method. In particular we focus on the fluctuations in the electronic energy level positions induced by interactions with the polar solvent. The simulation setup is shown in figure~\ref{6bp_h2o_scatreg}.
\begin{figure}[htbp]
\centerline{\epsfxsize=0.5\textwidth \epsffile{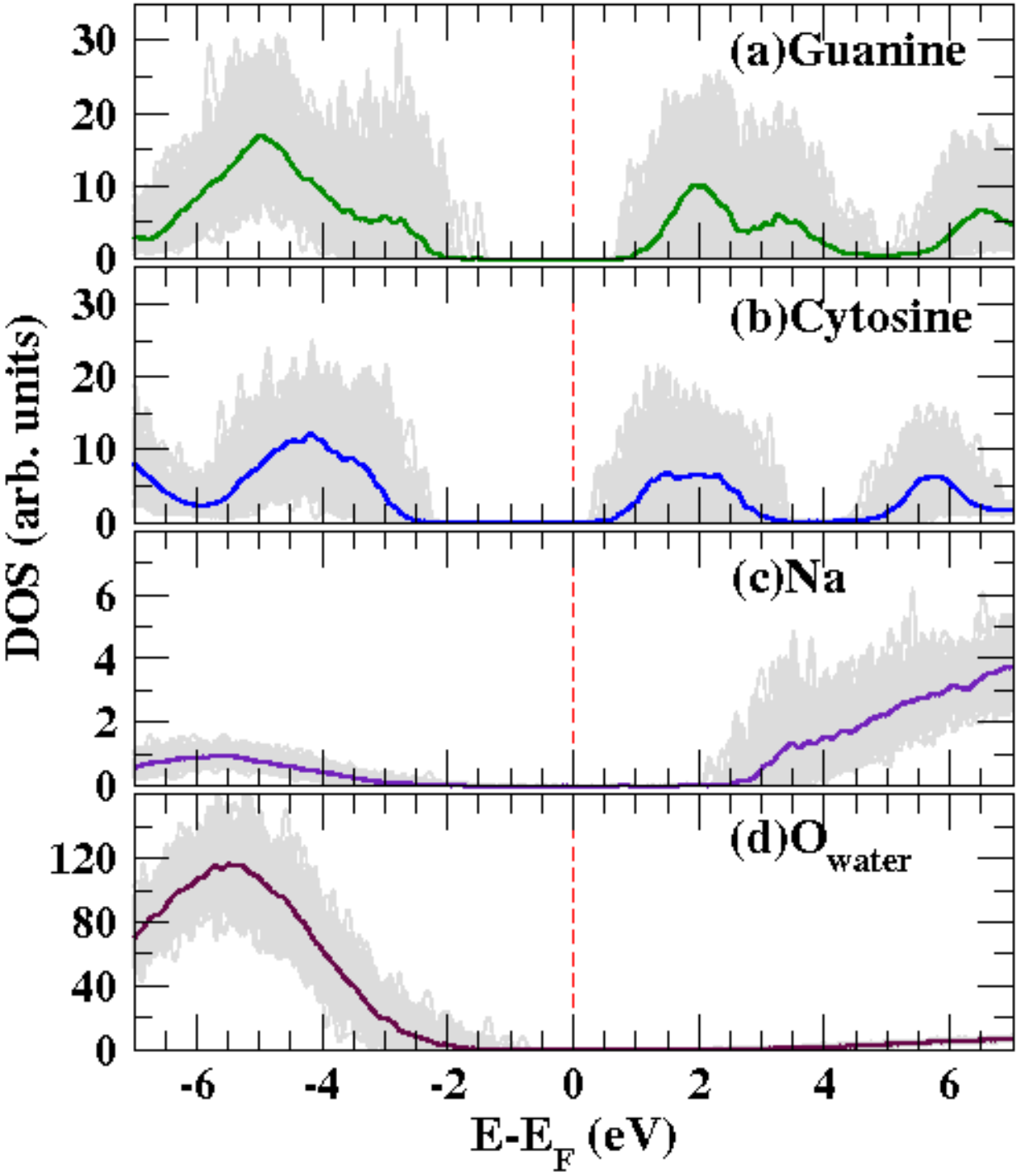}}
\caption{(Color online) Projected DOS for different components of the DNA and solvent system. The bold  dark lines show the PDOS averaged 
over the 100 MD snapshots. PDOS from all the individual snap-shots are shown in gray.}
\label{AvgDOS_all}
\end{figure}
\begin{figure}[htbp]
\centerline{\epsfxsize=0.5\textwidth \epsffile{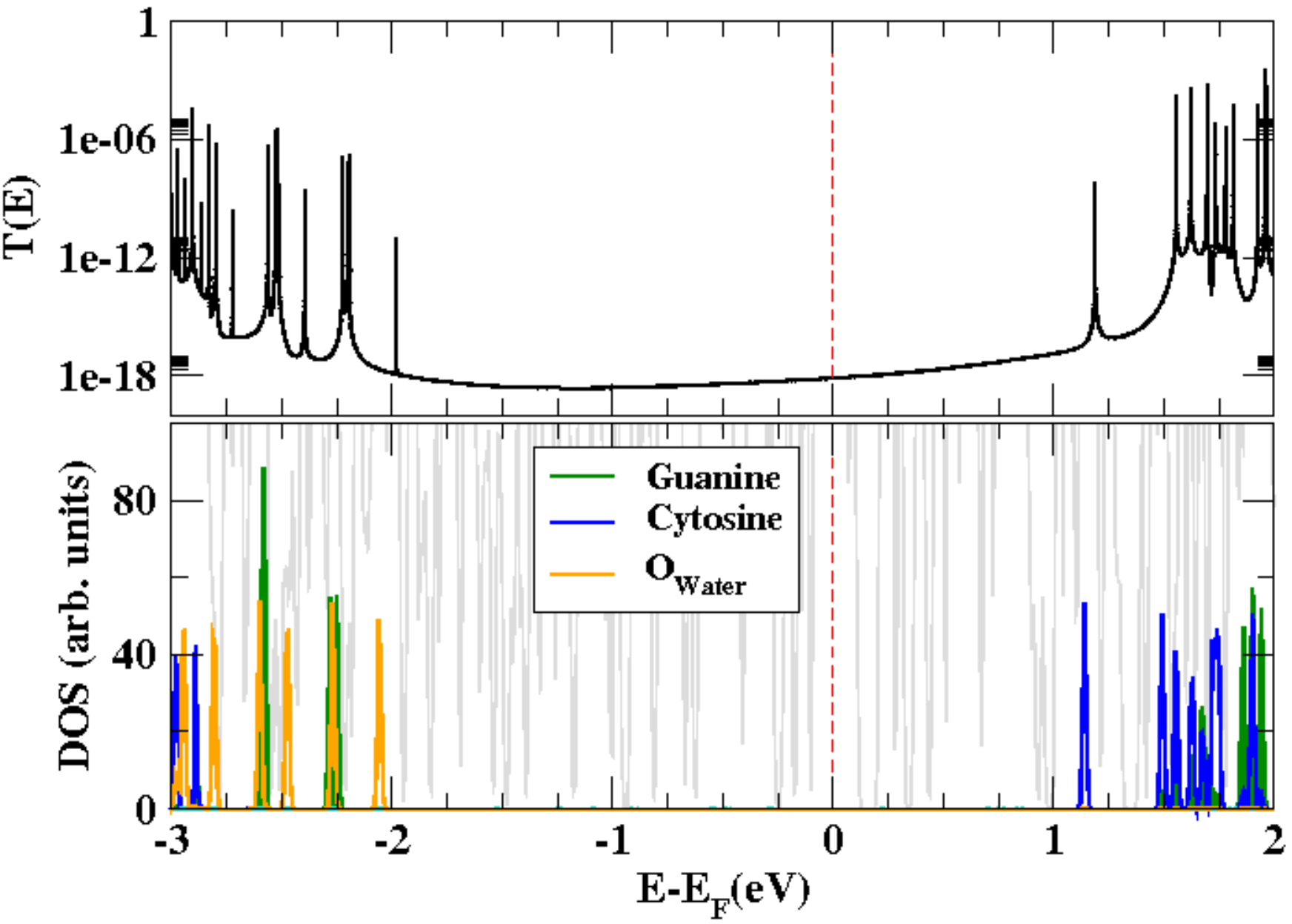}}
\caption{(Color online) Zero-bias transmission coefficient and corresponding DOS for one representative time snapshot of the DNA and 
solvent system taken from classical MD simulations. }
\label{sample_trcdosl}
\end{figure}
We consider a pGpC 6 base pair A-DNA duplex attached to Au electrodes in the C3$^\prime$G3$^\prime$ geometry as described earlier 
but now surrounded by a few layers of water molecules in the presence of Na$^{+}$ counter-ions. The
strategy adopted for investigating the effects of the solution on the electronic transport is the same reported recently in reference 
[\onlinecite{Ivan_bdt}]. It consists first in performing classical molecular dynamics (MD) simulations in order to describe the average 
geometrical arrangement of the solvent around the DNA, and then in evaluating the average transmission over a number of non-time-correlated configuration snapshots. 

In the dry condition the DNA backbone is neutral because each phosphate group is bonded to a hydrogen. In contrast in aqueous solution, 
the overall system neutrality is ensured by including two Na$^{+}$ counter-ions per base-pair. In order to describe the water solvation shell 
we perform empirical-potential MD simulations by using the {\sc namd2} package~\cite{namd}, parametrized with the standard additive CHARMM27 force field for nucleic acids~\cite{MacKerell1998}. For water we adopt the TIP3P model~\cite{Jorgensen1983} that is modified in the CHARMM force field and includes Lennard-Jones parameters for the hydrogen atoms as well as for oxygen~\cite{Reiher1985}. The interaction 
between the H2O molecules and gold is treated as a 12-6 Lennard-Jones potential as implemented in CHARMM force field, with parameters 
for Au ($\epsilon$=0.039 kcal/mol and $\sigma$=2.934~\AA) taken from the literature~\cite{namd_params}. Periodic boundary conditions 
are applied with a cutoff of 12~\AA for long-range interactions. The initial configuration of water molecules surrounding the pGpC 6 base 
pair DNA between the gold slabs is obtained using the VMD~\cite{Humphrey1996} solvation procedure. In order to maintain the size and 
shape of the cell constant, we perform simulations in the micro-canonical ensemble, with re-initialized velocities to 300~K for every 1000 
time-steps, each time-step 2~fs. 

Since our main objective is to examine the effects of solvation on the underlying electronic structure of A-DNA helices, we fix the atomic positions 
of the DNA strands in order to avoid conformational rearrangements. Thus only the water molecules and the Na$^{+}$ counter-ions are allowed 
to move during the MD simulation. After the initial equilibration of 1~ns we continue to run simulation for a total time of 20~ns. From the final 
configuration we extract the water solvation shell, the thickness of which is chosen so that both the back bone of the DNA helix as well as the 
counter-ions are adequately hydrated by water. In total, 211 water molecules and 12 Na$^{+}$ counter-ions are included in subsequent simulation 
at 300~K. The trajectory is recorded every 4~ps from the initial equilibration of 1~ns to a total simulation time of 20~ns.

We find that at equilibrium, the hydrated Na$^{+}$ counter-ions prefer to detach from the PO$_4^-$ groups of the DNA and remain suspended 
among the water molecules while the DNA back-bone is also strongly hydrated. Thus the interaction between the counter-ions and the DNA bases 
is strongly screened in aqueous solution.  

In order to extract a statistical picture of the energy level dynamics induced by molecule-solvent interactions, we select 100 uniformly spaced 
snapshots of the MD trajectory and calculate the electronic structure of each snapshot. The spacing between consecutive snapshots is large 
enough (120~ps) that the correlation between snapshots maybe assumed to be weak. In figure~\ref{6bp_h2o_scatreg}, which shows the planar 
average of electrostatic potential at zero bias plotted along the transport direction for all the snap shots considered, large fluctuations in the local 
electrostatic potential across different shapshots are apparent. These fluctuations are caused by the dynamics of the polar solvent molecules 
and act to locally gate the electronic energy levels of the bases along the DNA strand. As a result, the energy eigenvalues corresponding to 
molecular orbitals localized on the bases spread across a range of energies. 

In figure~\ref{HOLU_hist} we show histograms representing the number of times a certain molecular level is located in an energy range given 
by the bin width of the corresponding histogram. In figure~\ref{HOLU_hist}(a), the histograms for the HOMO and LUMO energies of the DNA 
strand are shown. Firstly we observe that the average HOMO level in the presence of the solvent is shifted down by approximately 1.5~eV relative to the HOMO of the molecule in the dry. The position of the average LUMO however changes little in comparison to the dry case. This is similar to what found for benzene molecules attached to gold via thiol groups \cite{Ivan_bdt}. Furthermore, the HOMO and LUMO energies fluctuate about the mean value with a standard deviation of 0.36~eV and 0.28~eV respectively. It is worth noting that although the peak of the histogram for the HOMO level in the presence of the solvent is $\sim$1.5~eV lower than the HOMO level in vacuum, energies near the right edge of the histogram are relevant for the observed I-V characteristics. The energy at which the average DOS for the occupied Guanine levels (see figure~\ref{AvgDOS_all}) starts to be substantial is seen to be $\sim$0.5~eV lower than the HOMO level in the dry. Furthermore, the HOMO levels discussed here correspond to neutral DNA strands and as such should correspond to vertical ionization potentials. If one also takes into account the possibility of solvent reorganization around the DNA~\cite{Basko, Mantz} in the case of a hole charge being present on the strand, the effective ionization potential can be lower.  Also shown in figure~\ref{HOLU_hist} are the histograms corresponding to energy levels localized on specific Guanine and Cytosine bases. The large amplitude of the fluctuations relative to the energy spacing between the molecular levels in the dry means that unlike in the static case, states localized on different bases can cross each other. In fact while in the dry the HOMO was always localized on the Guanine base at the 5$^\prime$ end of the DNA strand, the dynamic energy disorder induced by the solvent rearranges the order of the states in the HOMO manifold. This means that at a given moment in time (i.e. for a given water configuration) any of the energy levels associated to the six Guanine bases can be the HOMO. There is even the possibility that the instantaneous HOMO may be delocalized across several Guanine bases whose energy levels are close to resonant.  

\begin{table}[ht]
\begin{ruledtabular}
\begin{tabular}{lcccccc} 
&G1&G2&G3&G4&G5&G6\\
\hline
G1&-&63&62&69&64&79\\
G2&37&-&52&56&54&74\\
G3&38&48&-&64&59&75\\
G4&31&44&36&-&47&77\\
G5&36&46&41&53&-&75\\
G6&21&26&25&23&25&-\\
\end{tabular}
\caption{\label{Tab1}Table showing the number of time-snapshots in which the HO energy level localized on the Guanine base from the row 
entry is higher than the HO energy level localized on the Guanine base in the column entry.}
\end{ruledtabular}
\end{table}
In table~\ref{Tab1} we quantify the fluctuations between the states of the HOMO manifold. We list the number of snapshots in which an energy 
level localized on a given Guanine base is higher than all those localized on another bases. The Guanine bases are enumerated as G1 to G6 in sequential fashion starting from the 5$^\prime$ end of the strand. For example, it is seen that the energy level of the 
base G1 at the 5$^\prime$ end of the strand is higher than the energy level on the base G2 which is next to G1, only 63 percent of the time. 
Likewise, the energy level corresponding to the base G6 which is at the 3$^\prime$ end of the strand is higher than the level on the base G1 
about 21 percent of the time. From the the table it emerges that the energy levels associated to the terminal bases in the strand have a higher 
tendency of remaining at the edge of the HOMO manifold. Thus for instance the state corresponding to the G1 bases is more frequently the 
HOMO than any other level, and the state associated to the G6 bases is more frequently the deeper level in the manifold. However, all the other 
energy levels fluctuate quite considerably. 

The effects of such fluctuations on the electronic structure are discussed next. In figure~\ref{AvgDOS_all} we show the PDOS, averaged over 
the 100 snapshots considered, for different components of the system. In the same figure we display all the instantaneous DOS calculated 
at each time snapshot (gray lines merging in a shadow). We note that the net (average) effect of the solvent is to gate the occupied levels 
by shifting them to lower energies and effectively increasing the HOMO-LUMO gap. Note that the PDOS originating from the Na$^+$ counterions 
is largely confined to the unoccupied levels and located higher in energy than the primary peaks in the unoccupied DOS projected onto the 
Cytosine and Guanine bases. Thus the Na$^+$ counterions are not expected to play any significant role in either electron or hole transport 
through the DNA itself, with the exception of the fact that they contribute to the electrostatic disorder. 

In contrast, a large fraction of the PDOS for the solvent water molecules comes from the occupied levels and is peaked about 5~eV below the Fermi level. Importantly, however, a significant amount of the water DOS extends into the vicinity of the Guanine HOMO levels and therefore water molecules might also contribute to the electron transport across DNA. In figure~\ref{sample_trcdosl}  we show the transmission coefficient as a function of energy and the corresponding DOS for one time snapshot. The snapshot chosen is one in which the HO and LU molecular levels are very close to the average HOMO and LUMO values.  Qualitatively, there is very little difference between the transmission coefficients in the presence and in the absence of the solvent.  However, we observe that the presence of the water molecules enhances the tunneling transmission probability around zero bias compared to that of the dry situation which suggests that the decay of the wave-function at energies in the DNA bandgap is slower in water than it is in vacuum.
  
\section{Discussion and Conclusions}

The present work establishes a number of clear facts. Firstly, it shows that coherent tunneling across DNA strands attached to Au
electrodes with realistic bonding geometries is through the base-pair stack and not across the DNA backbone. This, however, is far from being band-like transport as speculated a long time ago on the basis of an analogy with aromatic salts \cite{Eley}. In fact, although an idealized homogeneous and periodic DNA chain possesses narrow conduction and valence bands, finite DNA fragments are dipolar molecules and such bands fragment into a manifold of extremely localized states that extend over one or two neighbouring bases. Furthermore, intra base-pair interactions between complementary bases are weak with the result that molecular states can be classified almost exclusively as belonging to one base or the other. In pGpC DNA strands in particular, the HOMO manifold has Guanine character while the LUMO has a Cytosine one. The coherent transport is then resonant tunneling-like across such localized states. Interestingly the transmission function is affected by the linker groups which attach the strand to the electrodes, even if those are connected to the backbone.

We then find that the $I$-$V$ characteristics are gapped with a conductance gap of the order of 2~Volt, in good agreement with controlled STS experiments. The transport is dominated by HOMO conduction and this essentially reflects the HOMO-LUMO gap and the alignment of the DNA molecular levels with the Fermi energy of the gold electrodes. Such an alignment has been carefully evaluated by calculating the IP and EA for the molecules with an accurate self-consistent DFT method and then by exporting the result to the transport calculations via the ASIC scheme. The agreement with experiments is however broken when one looks at the magnitude of the current at the onset voltage($\sim1$~Volt). The current is several orders of magnitude smaller than what measured in experiments even for strands shorter than those measured. This indicates that coherent tunneling is not the main mechanism for longitudinal transport in DNA even for short DNA fragments attached to electrodes.

The picture changes drastically when one considers the effects of a few water solvation layers on the transport. Water molecules create a dynamical dipolar field that re-arranges the relative energy positions of the individual states of both the HOMO and LUMO manifolds. The average HOMO manifold is shifted down in energy by about 1~eV while the LUMO remains roughly in the same position. In this situation then both HOMO and LUMO transport may become relevant although the position of the DFT-LUMO is too low within our current approach and so a conclusive statement on bipolar transport is not possible based on our results.  In addition we find that the presence of the water broadens further the HOMO and LUMO manifold, but more crucially it changes the position of the energy levels associated to any given individual base by as much as 1~eV. 

Based on all these {\it ab initio} results we can formulate a mechanism for electron transfer in DNA. This occurs by means of incoherent resonant tunneling through narrow energy levels spatially localized at the base-pairs. Levels are brought into resonant by the dynamical water dipolar field, which then acts as a main source of de-phasing. We note that the typical time-scale for the re-arrangement of the DNA energy level due to the solvent is 
of the order of a few femto-seconds i.e. it is considerably faster than the DNA
molecular vibration characteristic time. We then conclude that phonons play only a secondary role in DNA conduction, which in turn is determined by the dynamics of the solvent. Our calculations thus pave the way for the construction of quantitative models for DNA conduction based on a first principles Hamiltonian.

\section*{Acknowledgment}

This work is sponsored by Science Foundation of Ireland (Grants No. 07/RFP/PHYF235 and No. 07/IN.1/I945), by CRANN and by
the EU FP7 nanoDNAsequencing. Computational resources have been provided by the HEA IITAC project managed
by the Trinity Center for High Performance Computing and by ICHEC.

\small{

}

\end{document}